\documentclass[twocolumn,preprintnumbers,prl]{revtex4}
\usepackage{graphicx}
\usepackage{dcolumn}
\usepackage{bm}
\usepackage{color}

\begin{document}

\title{Time-Resolved Detection of Photon-Surface-Plasmon Coupling at the Single-Quanta Level}

\author{Chun-Yuan Cheng}
\email{hineg1212@hotmail.com}
\author{Jen-Hung Yang}
\author{Yen-Ju Chen}
\author{Chih-Hsiang Wu}
\author{Chia-Nan Wang}
\author{Chih-Sung Chuu}
\email{cschuu@phys.nthu.edu.tw}
\affiliation{Department of Physics, National Tsing Hua University, Hsinchu 30013, Taiwan\\
and Center for Quantum Technology, Hsinchu 30013, Taiwan}

\begin{abstract}
The interplay of nonclassical light and surface plasmons has attracted considerable attention due to fundamental interests and potential applications. To gain more insight into the quantum nature of the photon-surface-plasmon coupling, time-resolved detection of the interaction is invaluable. Here we demonstrate the time-resolved detection of photon-surface-plasmon coupling by exploiting single and entangled photons with long coherence time to excite single optical plasmons. We examine the nonclassical correlation between the single photons and single optical plasmons in such systems using the time-resolved Cauchy-Schwarz inequality. We also realize single optical plasmons with programmable temporal wavepacket by manipulating the waveform of incident single photons. The time-resolved detection and coherent control of single optical plasmons offer new opportunities to study and control the light-matter interaction at the nanoscale.
\end{abstract}

\maketitle

\section{I. Introduction} 

The interplay of the nonclassical light and surface plasmons has attracted considerable attention due to fundamental interests and potential applications. For example, various nonclassical properties such as the wave-particle duality \cite{Kolesov09,Dheur16}, antibunching~\cite{Kolesov09,Dheur16}, sub-Poissonian statistics~\cite{Akimov07}, Hong-Ou-Mandel interference \cite{Fakonas14,Cai14}, and entanglement \cite{Altewischer02,Fasel05,Dieleman17} were observed with the single or entangled optical plasmons. Potential applications in quantum information processing including the single photon source \cite{Chang06}, single-photon transistor \cite{Chang07}, waveguide quantum electrodynamics \cite{Kolchin11}, and quantum controlled-NOT gate \cite{Wang16} were also proposed and demonstrated. 

To gain more insight into the quantum nature of the photon-surface-plasmon coupling, time-resolved detection of the interaction is invaluable. In this work we exploit single and entangled photons with long coherence time to study the photon-surface-plasmon coupling in the quantum regime. The long coherence time of the incident photons not only enables the time-resolved detection but also allows the dynamical control of single optical plasmons' probability amplitude. As an example, we examine the nonclassicality of the entangled photons, where one photon transmits through a metallic nanohole array~\cite{Ebbesen98,Ghaemi98,Thio99,Popov00,Barnes03,Lalanne03,Martin-Moreno04,Chang05,Shin05,Genet07,McMahon07,Braun09,Gordon10,Tabatabaei15,Singh15,Balakrishnan16}, using the Cauchy-Schwarz inequality \cite{Clauser74} with a timing resolution finer than the coherence time of the entangled photons. As another example, we manipulate the temporal wavepacket of the single photons incident on the nanohole array. This in turn leads to the generation of single optical plasmons with a temporal wavepacket highly resembling that of the incident single photons, thus providing a novel way to control the single optical plasmons. Analogous to how manipulating single photons enriches quantum optics and photonic quantum technologies \cite{Cirac97,Inoue02,Gorshkov07,Kolchin08,Specht09,Belthangady10,Zhang12,Feng17,Wu19}, the coherent control of the single optical plasmons may open up new opportunities for quantum plasmonics and quantum information processing. For example, by designing the temporal wavefunctions of the guided single optical plasmons, the multipartite entanglement among the solid-state qubits can be created and dynamically controlled \cite{Chen11,Gonzalez-Ballestero14}. The absorption of single photons by a semiconductor quantum dot on a plasmonic waveguide may also be enhanced by using waveform-controlled single optical plasmon in a way similar to the enhanced absorption of single photons by atoms \cite{Zhang12}. The time-resolved detection and manipulation are suited to different coupling geometries or plasmonic nanostructures (for example, the plasmonic waveguides where the interaction time could be longer than the timing resolution), thus offering many opportunities to study and control the light-matter interaction at the nanoscale.

\begin{figure}[b]
\centering
\includegraphics[width=1 \linewidth]{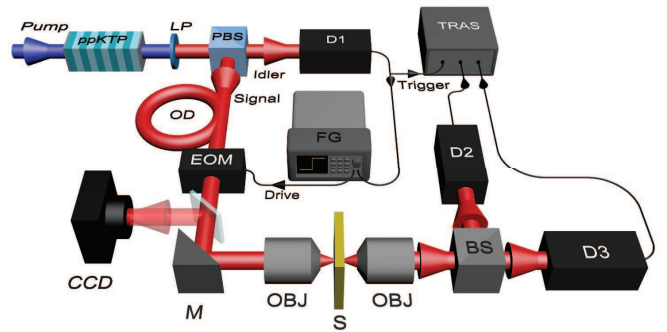}
\caption{\label{fig:1} Schematic of the experimental setup. TRAS: time-resolved analysis system, FG: function generator, EOM: electro-optic modulator, S: plasmonic nanostructure, OBJ: microscope objectives, LP: long-pass filter, PBS: polarizing beam splitter, OD: optical fiber, M: mirrors, BS: nonpolarizing beam splitter, and D1/D2/D3: single-photon detectors.}
\end{figure}

\section{II. Experimental setup} 

\begin{figure}[t]
\centering
\includegraphics[width=1 \linewidth]{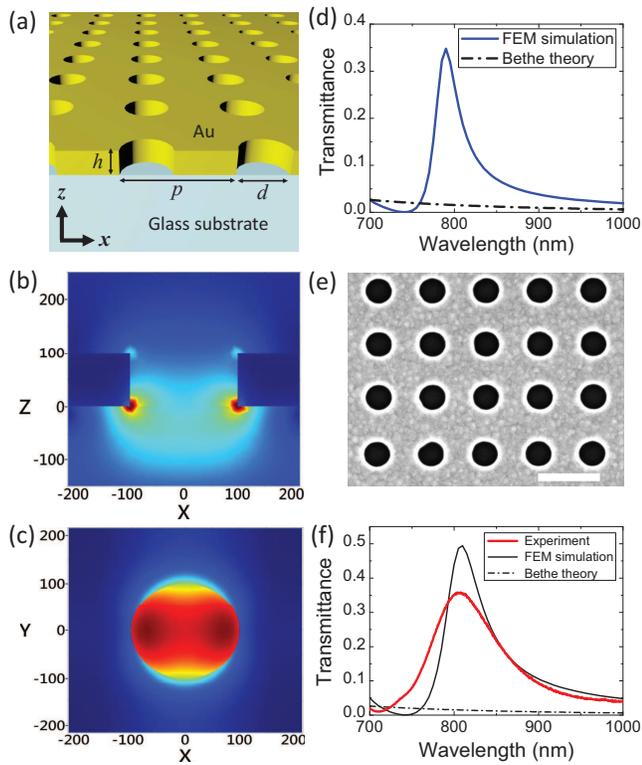}
\caption{\label{fig:2} (a) The plasmonic nanostructure consists of a metallic array of nanoholes. (b) Side view and (c) top view of the electric field near a nanohole. (d) The transmittance through the plasmonic nanostructure via the coupling with surface plasmon (solid curve) and the nanohole diffraction (dash-dot curve). (e) SEM image of the plasmonic nanostructure, where the scale bar is 500 nm. (f) Measured transmission spectrum (red solid curve) compared to the calculated transmission spectrum (black solid curve) and diffraction spectrum (black dash-dot curve). The simulation and theoretical curves in (b,c,f) are obtained by finite element method with the optical constants given in \cite{Johnson72}.}
\end{figure}

Fig.~\ref{fig:1} illustrates the experimental setup, where the resonant spontaneous parametric down-conversion in a monolithic cavity \cite{Chuu12,Wu17} is exploited to prepare single or entangled photons with long coherence time. The generating crystal, pumped by a 397.5-nm  frequency-doubled cw laser, emits time-energy entangled (signal and idler) photons near 795 nm with a paired rate of 2,000 s$^{-1}$ and a coherence time (FWHM width) of 50 ns. By spatially separating the orthogonally polarized signal and idler photons, the detection of the idler photons can be used to herald single photons with long coherence time in the signal channel. The detection of the idler photons also provides a precise time reference for the electro-optic modulator (which is driven by the function generator with programmable waveforms) to modulate the heralded single photons, so that the modulation starts simultaneously as the single photons (optically delayed by a 50-m-long optical fiber) arrives the modulator. To study the photon-surface-plasmon coupling, the signal photons are then tightly focused onto a plasmonic nanostructure, with the incident spot (diameter of 10 $\mu$m) carefully pre-examined by a charge-coupled device (CCD) camera to be free of defects. Finally, the single photons reemitted from the single optical plasmons are collected by a microscope objective for further analysis in order to study the properties of the single optical plasmons.

The plasmonic nanostructure is designed to optimize the conversion efficiency from the incident field to surface plasmon. It consists of a two-dimensional array of nanoholes on a gold film, which is illustrated in Fig.~\ref{fig:2}(a). The pitch, diameter, and film thickness are \textit{p} = 430 nm, \textit{d} = 200 nm and \textit{h} = 100 nm, respectively. As the incoming field, linearly polarized in the \textbf{\textit{x}}-direction, is incident along the \textbf{\textit{z}}-axis, the enhancement of the electric field can be seen near the edge of the nanohole in Fig.~\ref{fig:2}(b) (\textit{xz}-plane) and Fig.~\ref{fig:2}(c) (\textit{xy}-plane) as a result of the excited surface plasmon. The surface plasmon then leads to the field reemission on the other side of nanohole, with a transmission spectrum in Fig.~\ref{fig:2}(d) (blue solid curve) exhibiting resonance at the wavelength of our incident photon (795 nm). To ensure that the field transmission through the plasmonic nanostructure is dominated by the conversion between the incoming field and surface plasmon, the conversion efficiency is optimized so that the transmittance (0.31 at 795 nm) is greatly larger than that due to the nanohole diffraction (0.01 at 795 nm) calculated by the Bethe theory \cite{Bethe44} (black dash-dot curve).

\begin{table}[t]
\begin{tabular}{ccc}
		 \hline \hline
         & \ Incident photon\  & \ Reemitted photon\ \\   \hline 
     Unshaped & 0.019 $\pm$ 0.003   & 0.015 $\pm$ 0.003 \\  
     Shaped    & 0.019 $\pm$ 0.003   & 0.009 $\pm$ 0.003 \\  \hline \hline
\end{tabular}
\caption{Second-order quantum coherence function $g^{(2)}(0)$. The unshaped and shaped single photons have the waveforms of double exponential and exponential decay, respectively.}
\label{table:1}
\end{table}
 
Fig.~\ref{fig:2}(e) shows the scanning electron microscope (SEM) image of the plasmonic nanostructure, which is fabricated by sputtering 100-nm-thick gold film on a glass substrate followed by the focused ion beam (FIB) milling. The entire sample consists of 60 $\times$ 60 circular nanoholes with a dimension of 25.5 $\mu$m $\times$ 25.5 $\mu$m. The white ring surrounding the nanohole is the consequence of FIB's focusing angle, which results in tapered holes (angle of $17^{\rm o}$) with redshifted plasmonic resonance and increased transmittance. This can be seen in the calculated transmission spectrum (black solid curve), which takes into account the tapered holes, and the measured transmission spectrum (red solid curve) in Fig.~\ref{fig:2}(f), where the peak transmittance occurs at longer wavelength. In addition, since the gold film is polycrystalline, the surface of the gold film is not perfectly flat. The hole shape and size after the FIB milling thus vary slightly from hole to hole. This variation causes the transmission spectrum to broaden, with the measured peak transmittance (0.36) and full-width-at-half-maximum (FWHM) (96 nm) deviated from the calculated values (0.5 and 54 nm, respectively). Nevertheless, the wavelength of our single photon is still near the resonance peak with a measured transmittance of 0.34 much greater than that by diffraction (0.01, black dash-dot curve). To further verify the excitation of surface plasmon, we also study the polarization and angle dependence of the transmission spectrum. As the angle of incidence increases, we observe the splitting of the TM-mode spectrum into two peaks with increasing wavelength difference. In contrast, the spectra of the TE mode are independent of the angle of incidence. This angle dependence of the polarization-resolved transmission spectrum is another manifestation of the surface-plasmon-assisted transmission~\cite{Ebbesen98}.

\begin{figure}[t]
\centering
\includegraphics[width=1 \linewidth]{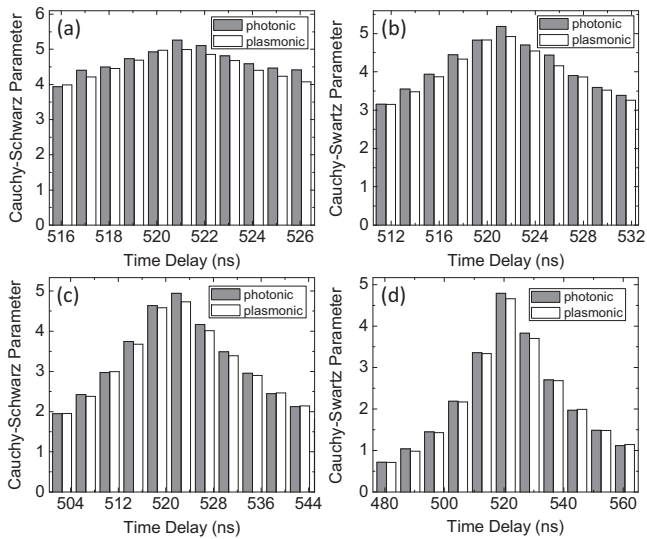}
\caption{\label{fig:3} Time-resolved detection of the nonclassical correlation between the single photons and single optical plasmons using the Cauchy-Schwarz inequality. The time bins used in (a-d) are 1, 2, 4, and 8 ns, respectively. The gray and white columns are the Cauchy-Schwarz parameters of the entangled photons with neither and one photon, respectively, transmitting through the metallic array of nanoholes.}
\end{figure}

\section{III. Time-resolved detection of nonclassicality} 

To confirm the generation of single optical plasmons from the signal photons, we verify the antibunching of the reemitted single photons in a Hanbury-Brown-Twiss-type experiment \cite{Hanbury-Brown56}. The reemitted photons are first incident on a beam splitter and the coincidence counts between the single-photon detectors at the two output ports, conditional on the detection of the idler photons, are then measured to calculate the second-order quantum coherence function $g^{(2)}(0) = p_{123}/p_{12}p_{13}$~\cite{Grangier86} at zero time delay, where $p_{123}$ is the joint probability of detecting one photon at both output ports and $p_{12}$ or $p_{13}$ is the probability of detecting one photon at each output port. Table~\ref{table:1} summarizes the $g^{(2)}(0)$ of the reemitted and incident photons, with the incident photons either unshaped or shaped. The single photons transformed back from the single optical plasmons not only preserve the single-photon nature ($g^{(2)}(0) < 0.5$) of the incident single photons, but also show a slightly smaller $g^{(2)}(0)$--a somewhat surprising result that may be contributed by the reduced noise (mostly the broadband fluorescence accompanied with the parametric down-conversion) due to the finite bandwidth of the plasmon resonance. 

With the signal photons converted to single optical plasmons on the metallic array of nanoholes, the nonclassicality of the entangled idler photon and optical plasmon can be examined by the nonclassical correlation of the idler photons and the reemitted photons using the Cauchy-Schwarz inequality \cite{Clauser74} $C(\tau) = g^2_{i,r}(\tau)/ g_{i,i}(0) g_{r,r}(0) \leq 1$, where $i$ and $r$ denote the idler photon and reemitted photon, respectively, $g_{i,r}(\tau)$ is the normalized cross-correlation function, $g_{i,i}(\tau)$ and $g_{r,r}(\tau)$ are the normalized auto-correlation functions, and $\tau$ is the time delay between the idler and reemitted photons. In the presence of the nonclasical correlation between the idler photon and optical plasmon (or the reemitted photon), which stems from their time-energy entanglement, the Cauchy-Schwarz inequality can be violated. The time-resolved tests of the inequality for different time bin sizes (1, 2, 4, and 8 ns) in Fig.~\ref{fig:3} show that the Cauchy-Schwarz parameter $C(\tau)$ and the nonclassicality varies with the probability amplitude of the incident entangled photons (whose coherence time is 50 ns) as a function of time.

\begin{figure}[t]
\centering
\includegraphics[width=0.8 \linewidth]{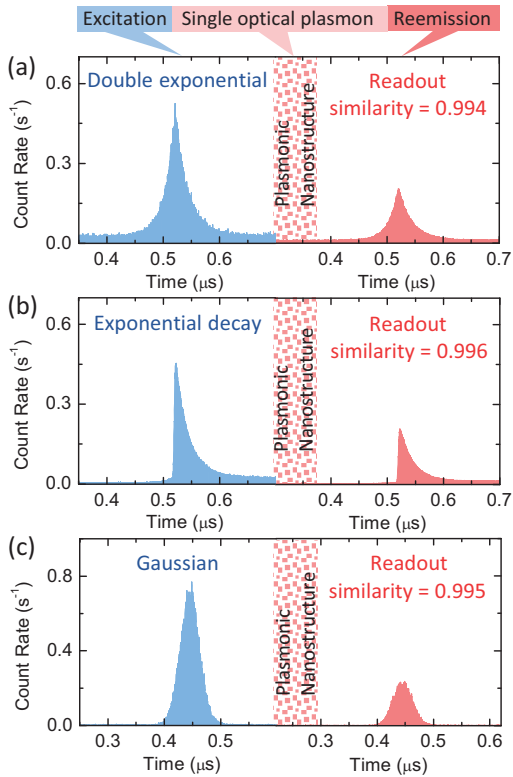}
\caption{\label{fig:4} The temporal wavepacket of the single optical plasmons is manipulated by shaping the incident single photons (left panels), which have the waveforms of (a) double exponential, (b) exponential decay, or (c) Gaussian distribution, and analyzed by the temporal wavepackets of the reemitted photons (right panels).}
\end{figure}

\section{IV. Coherent control of single optical plasmons}

We next demonstrate the viability to control the temporal wavepacket of the single optical plasmons. Fig.~\ref{fig:4}(a) (left panel) shows the unshaped temporal wavepacket of the incident signal photon as measured by the time-resolved intensity correlation between the signal and idler photons, which is proportional to the Glauber two-photon correlation function $G^{(2)}(\tau) = \langle a^{\dagger}_i (t+\tau) a^{\dagger}_s (t) a_s (t) a_i (t+\tau) \rangle$ with $a^{\dagger}(t)$ and $a(t)$ denoting the time-domain creation and annihilation operators, respectively, and $\tau$ being the time delay between the detection of the signal and idler photons. It has a waveform of double exponential distribution due to the resonant parametric down-conversion. This waveform, after the single photons excite the single optical plasmons, is imprinted onto the wavefunction of the single optical plasmons as evident by the waveform of the reemitted single photons in the right panel of Fig.~\ref{fig:4}(a). The conversion efficiency from the incident to reemitted single photons is 44\%. To see how resemble the two waveforms are, we calculate the cosine similarity (which measures the similarity between two non-zero data sets or vectors by computing the cosine of the ``angle'' between them) and obtain 0.994. The high fidelity of the waveform imprinting indicates that we can prepare single optical plasmons with programmable wavefunctions by simply controlling the waveform of the incident single photons. As an example, we demonstrate single optical plasmons with an exponential-decay wavepacket and a sharp front edge in Fig.~\ref{fig:4}(b), where the incident single photons are modulated by the Heaviside step function. The broad bandwidth of the plasmon resonance allows the faithful creation of a sharp edge in the wavecket of the single optical plasmon. As another example, we generate single optical plasmons with a Gaussian waveform of 40-ns FWHM width in Fig.~\ref{fig:4}(c). Here, the incident single photons are also shaped into the Gaussian waveform. In these examples, the cosine similarity are 0.996 (exponential-decay waveform) and 0.995 (Gaussian waveform), thus manifesting the high-fidelity waveform imprinting and the feasibility of generating single optical plasmons with programmable wavefunctions.

\begin{figure}[t]
\centering
\includegraphics[width=1 \linewidth]{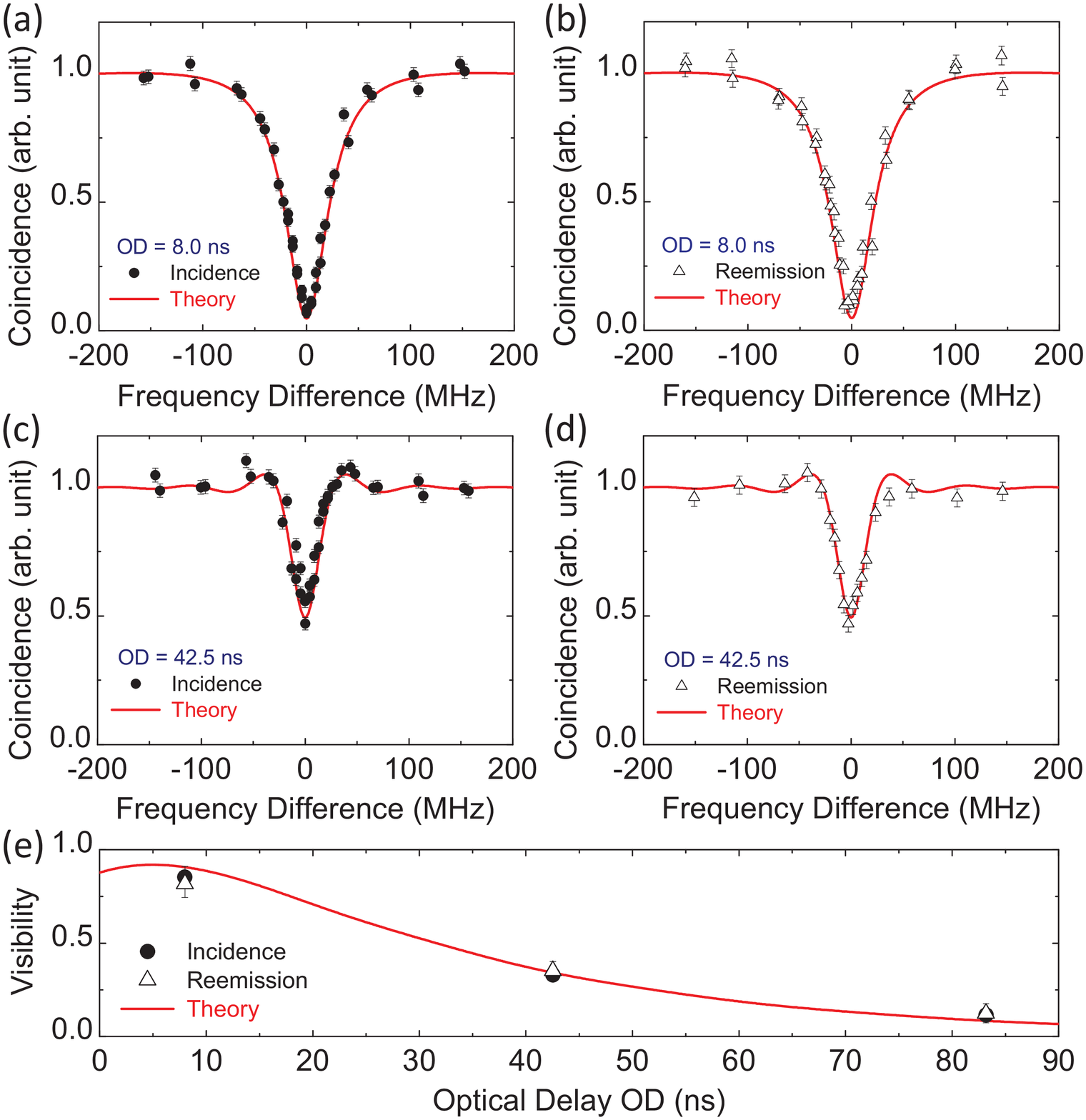}
\caption{\label{fig:5} Frequency-domain Hong-Ou-Mandel interference. The optical delays (ODs) are 8.0 ns in (a,b) and 42.5 ns in (c,d). (e) Coherence time of the single optical plasmons versus optical delays.}
\end{figure}

To gain more insight into the wavepacket transfer, we also carry out the frequency-domain Hong-Ou-Mandel (HOM) interference \cite{Hong87} between the reemitted signal photons and the idler photons, where the frequency difference between the signal and idler photons is adjusted by controlling the pump frequency and the crystal temperature. Figs.~\ref{fig:5} shows the coincidence counts between the output ports of the beam splitter for different optical delays between the signal and idler photons. Since indistinguishable photons always leave the beam splitter together, the coincidence counts increases with the frequency difference. Of importance, the visibilities and shapes of the two-photon interference after the wavepacket transfer [Figs.~\ref{fig:5}(b,d); divided by the transmittance through the sample] remain the same as those before the transfer [Figs.~\ref{fig:5}(a,c)] and agree with the theory (red curves), which computes the area of the temporal two-photon wavepacket existing through different output ports of the beam splitter. Moreover, as evident by the visibilities of the HOM interference for various optical delays in Fig.~\ref{fig:5}(e), the coherence time of the incident and reemitted single photons are identical within the statistical error and are in good agreement with the theory in the absence of wavepacket transfer (red curve). The nearly unchanged coherence time and waveforms (Fig.~\ref{fig:4}) of the reemitted single photons indicate that the spectrum of the incident single photons, which has a bandwidth of 4.5 MHz \cite{Chuu12,Wu17}, and the spectrum of the reemitted single photons are alike. This can also be seen from the similarity of the frequency-domain HOM interference, which is in essence the convolution of the spectrum and itself, of the incident and reemitted single photons (Fig.~\ref{fig:5}). By comparing their spectra to the same theoretical curve (red), we obtain cosine similarities larger than 0.996. As the incident photons are converted into surface-plasmon waves by the nanohole array and tunnel through the holes where the photons can go back and forth before reradiating as photons, the resemblance of the incident and reemitted single photons implies that coherent constructive interference \cite{Martin-Moreno01} may have been built up inside the holes.

\section{V. Conclusion} 

We have demonstrated the time-resolved detection and control of the photon-surface-plasmon coupling in the quantum regime, which are made possible by the long coherent time of the single and entangled photons. In particular, the nonclassical correlation between the incident and reemitted photons is tested by the time-resolved Cauchy-Schwarz inequality. By manipulating the waveform of the incident single photons, we also realize single optical plasmons with programmable temporal wavepacket. The time-resolved detection and coherent control of single optical plasmons are suited to different coupling geometries or plasmonic nanostructures. We note that more interesting properties of photon-surface-plasmon coupling may be studied with the method demonstrated in our work. For example, using a tunable filter of ultra-narrow bandwidth ($< 100$~kHz), one can explore the time-energy entanglement of the idler and reemitted photons by the separability criterion \cite{Duan00,Mancini02} or steering inequality \cite{Jones07,Wiseman07,Cavalcanti09}. Using an on-chip surface-plasmon detector \cite{Heeres09}, one can also study the correlation between the single photons and surface plasmons or surface plasmon polaritons. Our work thus offers new opportunities to study and control the light-matter interaction at the nanoscale, with potential applications including the controlled generation of multipartite entanglement \cite{Chen11,Gonzalez-Ballestero14} and the optimal absorption of single photons by quantum dots coupled to plasmonic waveguides. 

\section{Acknowledgments}
 
\begin{acknowledgments}
The authors would like to thank S. Gwo, J.S. Huang, Z.H. Lin and C.W. Cheng for their helpful discussion and experimental assistance. This work was supported by the Ministry of Science and Technology, Taiwan (107-2112-M-007-004-MY3, 107-2627-E-008-001 and 107-2745-M-007-001).
\end{acknowledgments}

\end{document}